# Dynamics of Heterogeneous Populations and Communities and Evolution of Distributions

Georgy P. Karev

Oak Ridge Institute for Science and Education (ORISE)
Bldg. 38A, Rm. 5N511N, 8600 Rockville Pike, Bethesda, MD 20894
karev@ncbi.nlm.nih.gov

**Abstract.** Most population models assume that individuals within a given population are identical, that is, the fundamental role of variation is ignored. Inhomogeneous models of populations and communities allow for birth and death rates to vary among individuals; recently, theorems of existence and asymptotic of solutions of such models were investigated. Here we develop another approach to modeling heterogeneous populations by reducing the model to the Cauchy problem for a special system of ODEs. As a result, the total population size and current distribution of the vector-parameter can be found in explicit analytical form or computed effectively. The developed approach is extended to the models of inhomogeneous communities.

## 1.     Statement of problems

The dynamic model of population $N(t)$ characterized by the growth factor $F(N,a)$ is given by the equation $dN/dt = NF(N,a)$ where the parameter $a$ is constant for all individuals in the population. Most growth models suppose that all individuals have identical attributes, in particular, identical rates of growth, death and birth. This assumption simplifies computation at the cost of realism. Recognition of heterogeneity can lead to unexpected and even counter-intuitive effects. Let us assume that each individual has its *own value* of the parameter $a$, which describes its invariable property (for example, a hereditary attribute or a specificity of the local habitat). The parameter remains unchanged for any given individual but varies from one individual to another, such that the population is non-uniform. Any changes of mean, variance and other characteristics of the parameter distribution with time are caused only by variation of the population structure. The theorems of existence and uniqueness of the solution of inhomogeneous models were proved in (Ackleh, 1998; Ackleh et al.,1999). Another approach to modeling heterogeneous populations was suggested in (Karev, 2000a,b).

The dependence of growth coefficient on a few (2-3) parameters is typical for most inhomogeneous models of population and biological communities. An opposite situation appears in problems of mathematical genetics, where a whole genome could be considered as a vector-parameter $a$ of a length $\sim 10^6$-$10^9$. Then $i=1,\ldots s$ is a number of genes ($s \sim 30,000$ for *Homo S*), $\varphi_i(a)$ is a measure of phenotype fitness of $i$-th gene and the total fitness is a weighted sum of all phenotype fitnesses.

Both situations could be studied in frameworks of unique general model. Let us suppose that each individual is characterized by its own value of the vector-parameter $a=(a_1,a_2,\ldots a_n)$ with the domain $A$. A set of all individuals with a given value of $a$ in the population is called $a$-group. Let $l(t,a)$ be a density of the number of individuals in the $a$-group at time $t$. Suppose that the reproduction rate of every *a*-group does not depend on other groups, but can depend on its value of the parameter *a* and the total size of the population *N*. Then we get the master model:

$$dl(t,\mathbf{a})/dt = F(N,\mathbf{a})\, l(t,\mathbf{a}), \tag{1}$$

$$N(t) = \int_A l(t,\mathbf{a})d\mathbf{a}, \quad l(0,\mathbf{a}) \text{ is given.}$$

We suppose here that the growth rate is of the form

$$F(N,\mathbf{a}) = f(N) + \sum_{i=1}^{n} g_i(N)\varphi_i(\mathbf{a}). \tag{2}$$

Let us denote $P(t,\mathbf{a})$ the current probability density function (pdf) of the vector-parameter $\mathbf{a}$, $P(t,\mathbf{a}) = l(t,\mathbf{a})/N(t)$. In model (1)-(2) we have a family of random variables $\varphi_i$, given initially on the probabilistic space $(A,P(0,\mathbf{a}))$; the model determines the temporary evolution of the pdf $P(t,\mathbf{a})$. The problem is to find these pdf and/or distributions of the random vector $(\varphi_1,\ldots \varphi_n)$ and the total population size $N(t)$ under the given initial parameter distribution. Although model (1)-(2) defines, in general, a very complex transformation of the probabilities $P(t,\mathbf{a})$, this problem can be solved effectively.

## 2. Basic Theorems

The main assertions about the master model are formulated in the following theorems 1-4. Denote $\vartheta=(\varphi_1, \varphi_2,\ldots \varphi_n)$ and $p(t,\vartheta)$ the pdf of the random vector $\vartheta$ at *t* moment, i.e. $p(t,x_1, \ldots x_n) = P(t,\varphi_1=x_1,\ldots \varphi_1=x_1)$. Let $\lambda=(\lambda_1,\ldots \lambda_n)$; denote

$$M(t;\lambda) = \int_A \exp(\sum_{i=1}^{n} \lambda_i \varphi_i(\mathbf{a})) P(t,\mathbf{a}) d\mathbf{a} = \tag{3}$$

$$\int_{\varphi(A)} \exp(\sum_{i=1}^{n} \lambda_i x_i)\, p(t, x_1,\ldots x_n) dx_1,\ldots dx_n$$

the moment generation function (mgf) of the pdf $p(t,\vartheta)$ of the random vector $\vartheta$. The mgf of the initial pdf of $\vartheta$, $M(0;\lambda)$, plays a fundamental role in the theory. In what follows we denote $E^t\phi = \int_A \phi(\mathbf{a}) P(t,\mathbf{a}) d\mathbf{a}$. Let us introduce *auxiliary variables* $q_i(t)$, $i=0,\ldots n$, as a solution of the system of differential equations:

$$dq_i(t)/dt = g_i(N^*(t)),\ q_i(0)=0 \text{ for } i=1,\ldots n, \tag{4}$$

$$dq_0(t)/dt = q_0(t) f(N^*(t)),\ q_0(0)=1,$$

$$N^*(t) = N(0)\, q_0(t) M(0,\mathbf{q}(t)) \tag{5}$$

where $\mathbf{q}(t) = (q_1(t),\ldots q_n(t))$.

Let $0 < T \leq \infty$ be the maximal value of *t* such that Cauchy problem (4), (5) has a unique solution in $[0,T)$. The existence of a global solution of this Cauchy problem critically depends on the initial pdf $P(0,\mathbf{a})$. For example, at $n=1$ and $\phi(a)=a$, the mgf $M(0, t)$ is defined for all $-\infty < t < \infty$ for the normal distribution, for $t < T = \inf (u > 0\colon q_0(u) = s)$ for the exponential distribution with the mean $1/s$, and only for $t\colon q_0(t) \leq 0$ for the log-normal distribution. The investigation of the existence of a global solution of Cauchy problem (4), (5) for a given initial distribution and functions $f$, $g_i$ can be a non-trivial problem. The following main theorem reduces model (1)-(2) to the Cauchy problem.

**Theorem 1.** *Let Cauchy problem* (4), (5) *has a unique solution* $\{q_i(t)\}$ *at* $t \in [0,T)$. *Then the functions*

$$l(t,\boldsymbol{a}) = l(0,\boldsymbol{a})\, q_0(t)\exp[\sum_{i=1}^{n} \varphi_i(\boldsymbol{a})\, q_i(t)], \qquad (6)$$

$N(t) = N(0)q_0(t)M(0,\boldsymbol{q}(t))$
*satisfy system* (1)-(2) *at* $t \in [0,T)$.
*Conversely, if* $l(t,\boldsymbol{a})$ *and* $N(t) = \int_A l(t,\boldsymbol{a})\mathrm{d}\boldsymbol{a}$ *satisfy system* (1)-(2) *at* $t \in [0,T)$, *then Cauchy problem* (4) *has a solution* $\{q_i(t)\}$ *at* $t \in [0,T)$ *and the given functions* $l(t,\boldsymbol{a})$, $N(t)$ *can be written in the form* (6) *at* $t \in [0,T)$.

Let us emphasize that the inhomogeneous model (1)-(2) determines not only dynamics of the total size $N(t)$ of the population, but also the evolution of the parameter distribution $P(t,\boldsymbol{a})$. This aspect of the inhomogeneous models is essentially new compared to "regular", homogeneous, dynamical models. Sometimes, the evolution of the parameter distribution is of principal concern. The following important corollary of Theorem 1 explains the evolution of the composition of the inhomogeneous population.

**Corollary 1.** $l(t,\boldsymbol{a}_1)/l(t,\boldsymbol{a}_2) = l(0,\boldsymbol{a}_1)/l(0,\boldsymbol{a}_2)\, \exp(\int_0^t [F(N(s),\boldsymbol{a}_1) - F(N(s),\boldsymbol{a}_2)]\mathrm{d}s.$

This implies the *Haldane dynamic optimality principle* in the following form.

Denote $k(\boldsymbol{a}) = \lim[1/t \int_0^t F(N(s),\boldsymbol{a})]$ the mean reproduction rate of an $\boldsymbol{a}$-group.

Then, the evolution of a heterogeneous population leads to an (exponentially fast) replacement of individuals with smaller values of $k(\boldsymbol{a}_1)$ by those with greater values of $k(\boldsymbol{a}_2)$, even though the fraction of the latter in the initial distribution was arbitrarily small.

To investigate the evolution of the parameter distribution, we need the following

**Lemma 1**. *Let* $M(0,\boldsymbol{\lambda})$ *be the mgf at the initial moment* $t=0$ *and the Cauchy problem* (4), (5) *has a unique solution* $\{q_i(t)\}$ *at* $t \in [0,T)$. *Then for all* $t \in [0,T)$
$M(t;\boldsymbol{\lambda}) = M(0;\boldsymbol{\lambda}+\boldsymbol{q}(t))/M(0;\boldsymbol{q}(t))$.

Theorem 2 describes the dynamics of distributions of the parameter.
**Theorem 2.** *Under conditions of Theorem* 1,
(i) *The current pdf* $P(t,\boldsymbol{a})$ *is determined by the formula*

$$P(t,\boldsymbol{a}) = P(0,\boldsymbol{a})\, \exp[\sum_{i=1}^{n} \varphi_i(\boldsymbol{a})\, q_i(t)] / M(0;\boldsymbol{q}(t));$$

(ii) *The pdf* $P(t,\boldsymbol{a})$ *solves the equation*

$$\mathrm{d}P(t,\boldsymbol{a})/\mathrm{d}t = P(t,\boldsymbol{a})\, [\sum_{i=1}^{n} g_i(N(t))(\varphi_i(\boldsymbol{a}) - E_t\varphi_i)].$$

The following theorem describes the dynamics of the population sizes and the main statistical characteristics of model (1)-(2). Let us denote $K(t;i,k)$ the covariance of the random variables $\varphi_i(\boldsymbol{a})$, $\varphi_k(\boldsymbol{a})$ over the pdf $P(t,a)$,
$K(t;i,k) = E^t\{[\varphi_k(\boldsymbol{a})-E^t\varphi_k][\varphi_i(\boldsymbol{a})-E^t\varphi_i]\}$. Denoted also $\partial_i F(\boldsymbol{x}) = \partial F(x_1,\ldots x_n)/\partial x_i$.
**Theorem 3.** *Under conditions of Theorem* 1,
(i) *the current population size* $N(t)$ *satisfies the equation*

$$\mathrm{d}N/\mathrm{d}t = N\,[f(N) + \sum_{i=1}^{n} q_i(N)\, E^t\varphi_i] = N\, E^t[F]; \qquad (2.4)$$

(ii) *the current mean values* $E^t\varphi_i$ *are determined by the formula*

$E^t\varphi_i = [\partial_i M(0;\boldsymbol{q}(t))] / M(0;\boldsymbol{q}(t))$ (2.5)

and satisfy the equation $d(E^t\varphi_i)/dt = \sum_{k=1}^{n} g_k(N)K(t;i,k)$;

(iii) *the current covariance $K(t;i,k)$ can be computed using the formula*
$K(t;i,k) = [\partial^2_{ik} M(0;\boldsymbol{q}(t)) - \partial_i M(0;\boldsymbol{q}(t)) \partial_k M(0;\boldsymbol{q}(t))]/ M(0;\boldsymbol{q}(t))$. (2.6)

**Corollary 2**. *The mean fitness $E^t[F(N(t), \boldsymbol{a})]$ satisfies the equation*
$dE^t[F]/dt = (\partial E^t[F]/\partial N) E^t[F] N + \boldsymbol{g}^T \boldsymbol{K}(t) \boldsymbol{g} = (\partial (E^t[F])^2/\partial N) N/2 + Var^t F$
*where $\boldsymbol{g}^T$ is a transposed vector $\boldsymbol{g} = (g_1(N),\ldots g_n(N))$, $\boldsymbol{K}(t) = \{K(t;i,k)\}$ is the covariance matrix of the r. v. $\{\varphi_i, i=1,..n\}$ and $Var^t F$ is the variance of $F(N(t),\boldsymbol{a})$ at t moment*.

Theorem 2 gives the methods of computation of the current pdf for master model (1)-(2) depending on the initial pdf. Theorem 3 describes the dynamics of the main statistic characteristics of the distribution of the vector $\vartheta=(\varphi_1, \varphi_2,\ldots \varphi_n)$. As an important particular case, let us consider the growth coefficient $F(N,\boldsymbol{a})= f(N)+ \varphi(\boldsymbol{a})g(N)$. The evolution of the distribution of $\varphi(\boldsymbol{a})$ can be investigated analytically for many cases of the initial distribution. The model is of the form
$dl(t,\boldsymbol{a})/dt = [f(N)+ \varphi(\boldsymbol{a})g(N)] l(t,\boldsymbol{a})$, (2.7)
$N(t) = \int_A l(t,\boldsymbol{a})d\boldsymbol{a}$, $P(t,\boldsymbol{a})=l(t,\boldsymbol{a})/N(t)$.

The system for the auxiliary variables is
$dq_0(t)/dt = q_0(t) f(N^*(t))$, $q_0(0)=1$, (2.8)
$dq_1(t)/dt = g(N^*(t))$, $q_1(0)=0$,
where $N^*(t)= N(0) q_0(t)M(0, q_1(t))$.

*(i)* **Theorem 4.** *Let Cauchy problem* (2.8) *have a unique solution $\{q_0(t), q_1(t)\}$ at $t\in[0,T]$. Let us assume that in the initial time moment the r.v. $\varphi(\boldsymbol{a})$ has*
*(ii)* (i) *normal distribution with a mean m and variance $\sigma^2$; then the pdf $p(t,\varphi)$ is also normal at any $t\in[0,T]$ with the mean $E^t\varphi = m+\sigma^2 q_1(t)$ and with the same variance $\sigma^2$;*
(ii) *Poisson distribution with a mean m; then the pdf $p(t,\varphi)$ is also Poissonian at any $t\in[0,T]$ with the mean $E^t\varphi = m \exp(q_1(t))$;*
(iii) *$\Gamma$-distribution with the coefficients k, s, $\eta$:*
$p(0,\varphi) = s(\varphi - \eta)^{k-1}\exp[-(\varphi - \eta)s]/\Gamma(k)$, (2.9)
*where $s>0$, $k>0$, $-\infty < \eta < \infty$ and $a \geq \eta$; $\Gamma(k)$ is the $\Gamma$-function.*
*Define $T^* = \inf(t\in[0,T]: q_1(t) = s)$, if such t exists, otherwise $T^* =T$. Then the pdf $p(t,\varphi)$ is also $\Gamma$-distribution at any time moment $t<T^*$ with coefficients $k$, $s- q_1(t)$, and $\eta$ such that $E^t\varphi = \eta + k/(s - q_1(t))$, $\sigma(t)^2 = k/(s - q_1(t))^2$.*

An important conclusion can be drawn relying on the last formulae: even arbitrary but non-zero variance of the initial $\Gamma$-distribution of the growth factor gives rise to the model «blowing up», i.e., the population size and also the mean and variance of the population distribution tend to infinity in finite time.
The evolution of other initial pdf can be investigated in the same way.
**Example**. The growth of the world population $N$ over hundreds of years, up to ~1990, was described with high accuracy by the hyperbolic law $N(t)=C/(T-t)^k$ where $C\approx 2*10^{11}$, $T\approx 2025$, $k\approx 1$ that predicts a demographic explosion at the time $T$ (Forster, 1960). This formula solves the quadratic growth model $dN/dt = N^2/C$, in which the individual reproduction rate is proportional to the total human population (apparently, this relationship makes no "biological" sense). The above results (see Theorem 4, iii)) show that the hyperbola is implied also by the Malthusian inhomogeneous model with a gamma-distributed reproduction rate. When the reproduction rate in the model is bounded, the result is that $N(t)$ is finite, even though indefinitely increasing, for all $t$. Therefore, the "demographic explosion" is a corollary of the obviously unrealistic assumption (incorporated implicitly into quadratic growth model) that the individual birth coefficient may take unlimitedly large values with *non-zero* probabilities.

The subsequent transition from the Malthusian model to the inhomogeneous logistic model (see (Karev, 2005) for details) show a notable phenomenon, a transition from protracted hyperbolical growth (the phase of "hyper-exponential" development) to the brief transitional phase of exponential growth and, subsequently, to stabilization. We could conclude that the hyperbolic growth of the humankind was not an exclusive phenomenon but obeyed the same laws as any heterogeneous biological population.

## 3. Evolution of the distributions of the vector $\vartheta$

We will suppose again that the Cauchy problem (4), (5) has a solution $(q_0(t),\ldots q_n(t))$ in $[0,T)$.

**Theorem 5.** *Let the random variables $\{\varphi_i, i=1,\ldots n\}$ be independent at the initial instant and have the initial pdf $p_i(0,x_i)$, so that $\boldsymbol{p}(0,x_1,\ldots x_n)= \prod_{i=1}^{n} p_i(0,x_i)$ and*

$M(0,\lambda_1,\ldots \lambda_n) = \prod_{i=1}^{n} M_i(0;\lambda_i)$ *where $M_i(0,\lambda_i)$ is the mgf of $p_i(0,x_i)$. Then for any $t \in [0,T)$ the r. v.-s $\varphi_i$ are independent, the mgf $M_i(t;\lambda_i)=M_i(0;\lambda_i+q_i(t))/M_i(0;q_i(t))$ and $M(t,\lambda_1,\ldots \lambda_n) = \prod_{i=1}^{n} M_i(0;\lambda_i+q_i(t))/M_i(0;q_i(t))$.*

The evolution of the pdf of the random vector $\vartheta=(\varphi_1, \varphi_2,\ldots \varphi_n)$ in the general case of correlated random variables $\{\varphi_i, i=1,\ldots n\}$ is of great practical interest because it helps to explore the dynamics of inhomogeneous population (1)-(2) depending on correlations between the random variables $\varphi_i(\boldsymbol{a})$.

**Definition**. *A class S of probability distributions of the random vector $\vartheta=(\varphi_1,\ldots \varphi_n)$ is called invariant with respect to model* (1)-(2) *in the interval* $[0, T)$, *if $\boldsymbol{p}(0,\vartheta)\in S \Rightarrow \boldsymbol{p}(t,\vartheta) \in S$ for all $t <T$.*

Let **MS** be the class of moment-generating functions for distributions from the class *S*. The criterion of invariance immediately follows from Lemma 1.

**Invariance criterion**. A class *S* of pdf is invariant in $[0, T)$ with respect to model (1)-(2) if and only if
$M(0,\boldsymbol{\lambda})\in \boldsymbol{MS} \Rightarrow M(0,\boldsymbol{\lambda}+\boldsymbol{q}(t))/M(0,\boldsymbol{q}(t)))\in \boldsymbol{MS}$ for all $t<T$.

We can prove with the help of this criterion that many important distributions are invariant with respect to model (1)-(2) in the interval $[0, T)$. Let us recall some definitions. A random vector $\boldsymbol{X}=(X_1,\ldots X_n)$ has multivariate normal distribution with the mean $E\boldsymbol{X}=\boldsymbol{m}=(m_1,\ldots m_n)$ and covariance matrix $\boldsymbol{K}=\{c_{ij}\}$, $c_{ij}=\mathrm{cov}(X_i,X_j)$ if its mgf is $E[\exp(\boldsymbol{\lambda}^T \boldsymbol{X})] =\exp[\boldsymbol{\lambda}^T\boldsymbol{m} +1/2\boldsymbol{\lambda}^T\boldsymbol{C\lambda}]$ (see Kotz, p.108).

A random vector $X=(X_1,\ldots X_n)$ has a multivariate polynomial distribution with parameters $(k; p_1,\ldots p_n)$, if $P\{X_1=m_1,\ldots X_n=m_n\}=\dfrac{k!}{m_1!\ldots m_n!}p_1^{m_1}\ldots p_n^{m_n}$ for $\sum_{i=1}^{n} m_i = k$. The mgf of the polynomial distribution is $M(\lambda)=(\sum_{i=1}^{n} p_i\exp(\lambda_i))^k$.

A general class of *multivariate natural exponential distribution* is most interesting for applications; this class includes multivariate polynomial, normal, and Wishart distributions as particular cases. A random vector $X=(X_1,\ldots X_n)$ has multivariate natural exponential distribution (NED) with parameters $\boldsymbol{\theta}=(\theta_1, \theta_2,\ldots \theta_n)$ with respect to the positive measure $\nu$ on $R^n$ if its joint density function is of the form $f_\theta(X)=h(X)\exp[X^T\boldsymbol{\theta}-s(\boldsymbol{\theta})]$ where $s(\boldsymbol{\theta})$ is a function on parameters (see [Kotz, ch. 44, 54]). The mgf of NED is $E[\exp(\lambda^T X)]=\exp[s(\boldsymbol{\theta}+\lambda)-s(\boldsymbol{\theta})]$.

**Theorem 6**. *Under conditions of Theorem* 1, *let us assume that in the initial time moment the random vector* $\vartheta=(\varphi_1, \varphi_2,\ldots \varphi_n)$ *has*

i) *multivariate normal distribution with the mean vector* $\boldsymbol{m}(0)$ *and covariance matrix* $\boldsymbol{K}=(c_{ik})$. *Then at any moment* $t<T$ *the vector* $\vartheta$ *also has the multivariate normal distribution with the same covariance matrix* $\boldsymbol{K}$ *and the mean vector* $\boldsymbol{m}(t)$ *with components* $m_i(t)= m_i(0)+1/2 \sum_{k=1}^{n} [c_{ik} + c_{ki}] q_k(t)$;

ii) *multivariate polynomial distribution. Then at any moment* $t<T$ *the vector-parameter* $\boldsymbol{a}$ *has the multivariate polynomial distribution with parameters* $(k; p_1(t),\ldots p_n(t))$, *where* $p_i(t)= p_i / \sum_{j=1}^{n} p_j\exp(q_j(t))$;

iii) *multivariate natural exponential distribution on* $R^n$ *with parameters* $\boldsymbol{\theta}=(\theta_1,\ldots \theta_n)$. *Then at any moment* $t<T$ *the vector* $\vartheta$ *also has the multivariate NED with the parameters* $\boldsymbol{\theta}+\boldsymbol{q}(t)$ *and the moment generating function* $\exp[s(\boldsymbol{\theta}+\lambda+\boldsymbol{q}(t))-s(\boldsymbol{\theta}+\boldsymbol{q}(t))]$.

*Remark*. On the contrary, the multivariate uniform distribution in a rectangle is not invariant (and turns at $t>0$ into the multivariate truncated exponential distribution in this rectangle).

## 4. Inhomogeneous models of biological communities

In the model of inhomogeneous population (1)-(2) we supposed that the growth factor $F(N,\boldsymbol{a})$ of $\boldsymbol{a}$-group does not depend on the sizes of other groups. This assumption may be weakened. Let the whole population be subdivided into $n$ subpopulations $N_1, \ldots N_m$; all individuals with a given value of $\boldsymbol{a}$ in the $j$-th population is called a $j$-th $\boldsymbol{a}$-group. Let $l_j(t,\boldsymbol{a})$ be the number of individuals in the $j$-th $\boldsymbol{a}$-group at time $t$. Suppose now that the reproduction rate of every $\boldsymbol{a}$-group does not depend on the size of other groups in the community but can depend on the corresponding value of the parameter vector $\boldsymbol{a}$ and the total sizes of subpopulations, $(N_1, \ldots N_m) = \boldsymbol{N}$. It means that the growth rate of every $\boldsymbol{a}$-group in $j$-th subpopulation is $F_j(N_1, \ldots N_m,\boldsymbol{a})= F_j(\boldsymbol{N},\boldsymbol{a})$; we assume that

$$F_j(\boldsymbol{N},\boldsymbol{a}) = f_j(\boldsymbol{N}) + \sum_{i=1}^{n} g_{ji}(\boldsymbol{N})\varphi_i(\boldsymbol{a}). \qquad (4.1)$$

Then we have the following model:

$dl_j(t,a)/dt = l_j(t,a) F_j(N,a),$ (4.2)

$N_j(t) = \int_A l_j(t,a)da.$

The dynamics of a community of $m$ interacting populations $N_1,...N_m$ with distributed parameters is described by the same model (4.1), (4.2). The theory developed for the model of inhomogeneous population can be extended to the case of inhomogeneous community. Let us present the generalization of main theorems for model (4.1), (4.2).

Let $P_j(t,a) = l_j(t,a)/N_j(t)$ be the current pdf of the parameter $a$ in the $j$-th population at $t$ instant; the initial pdf $P_j(0,a)$ is given. Let $M_j(t;\lambda) = \int_A \exp(\sum_{i=1}^n \lambda_i \varphi_i(a))P_j(t,a)da$; introduce *auxiliary variables* $q_{ji}(t)$, $j=1,...m$, $i=0,...n$ as a solution of the system of ordinary differential equations:

$dq_{ji}(t)/dt = g_{ji}(N^*(t)), q_{ji}(0)=0$ for $i=1,...n$, (4.3)

$dq_{j0}(t)/dt = q_{j0}(t) f_j(N^*(t)), q_{j0}(0)=1$

where $N^*_j(t) = N_j(0) q_{j0}(t) M_j(0; q_{j1}(t),... q_{jn}(t)) = N_j(0)q_{j0}(t)M_j(0;\mathbf{q}_j(t))$. (4.4)

**Theorem 7.** *Let $0<T\leq\infty$ be the maximal value of $t$ such that the Cauchy problem (4.3), (4.4) has a solution in $[0,T)$. Then the functions*

$l_j(t,a) = l_j(0,a) q_{j0}(t)\exp[\sum_{i=1}^n q_{ji}(t) \varphi_i(a)],$ (4.5)

$N_j(t) = N_j(0) q_{j0}(t)M_j(0; \mathbf{q}_j(t))$

*satisfy system (4.1), (4.2) at $t\in[0,T)$.*

*Conversely, if $l_j(t,a)$ and $N_j(t) = \int_A l_j(t,a)da$ satisfy system (4.1), (4.2) at $t\in[0,T)$, then Caushy problem (4.3) has a solution $\{q_{ji}(t)\}$ at $t\in[0,T)$ and the given functions $l_j(t,a)$, $N_j(t)$ can be written in the form (4.5) at $t\in[0,T)$.*

**Theorem 8.** *Under conditions of Theorem 7,*
*(i) The current pdf $P_j(t,a)$ are determined by the formula*

$P_j(t,a) = P_j(0,a) \exp[\sum_{i=1}^n \varphi_i(a) q_{ji}(t)] / M_j(0; \mathbf{q}_j(t))$ (4.6)

*and solve the equations*

$dP_j(t,a)/dt = P_j(t,a) [\sum_{i=1}^n g_{ji}(N(t))(\varphi_i(a) - E^t_j\varphi_i];$ (4.7)

*(ii) the current population sizes $N_j(t)$ satisfy the equation*

$dN_j/dt = N_j [f_j(N) + \sum_{i=1}^n q_{ji}(N) E^t_j\varphi_i] = N_j E^t_j F_j;$ (4.8)

*(iii) the current mean values $E^t_j\varphi_i$ are determined by the formula*
$E^t_j\varphi_i = [\partial_i M_j(0;\mathbf{q}_j(t))] / M_j(0;\mathbf{q}_j(t))$ (4.9)
*and satisfy the equation*

$d(E^t_j\varphi_i)/dt = \sum_{k=1}^n g_{jk}(N) K_j(t;i,k)$ (4.10)

where $K_j(t;i,k)$ is the correlation moment $K_j(t;i,k) = E^t_j\{[\varphi_k(a) - E^t\varphi_k] [\varphi_i(a) - E^t\varphi_i]\}$.

(iv) $dE'_j[F_j]/dt = \sum_{k=1}^{n} (\partial E'_j[F_j]/\partial N_k) E'_k [F_k] N_k + \mathbf{g}_j^T \mathbf{K}_j(t) \mathbf{g}_j$

where $\mathbf{g}_j^T$ is a transposed vector $\mathbf{g}_j=(g_{j1}(N),\ldots g_{jn}(N))$ and $\mathbf{K}_j(t)=\{K_j(t;i,k)\}$.

## 5. Proofs of the main theorems

*Proof of Theorem 1*
Let $\{q_i(t)\}$ be a solution of Cauchy problem (4) at $t\in[0,T)$; denote for instant
$l^*(t,\mathbf{a}) = l^*(0,\mathbf{a}) q_0(t) \exp[\sum_{i=1}^{n} \varphi_i(\mathbf{a}) q_i(t)]$. According to system (4),

$dl^*(t,\mathbf{a})/dt = l^*(0,\mathbf{a})[dq_0(t)/dt \exp[\sum_{i=1}^{n} \varphi_i(\mathbf{a}) q_i(t)] + l^*(t,\mathbf{a}) \sum_{i=1}^{n} \varphi_i(\mathbf{a}) dq_i(t)/dt] =$

$l^*(t,\mathbf{a})[f(N^*(t)) + \sum_{i=1}^{n} \varphi_i(\mathbf{a}) g_i(N^*(t))] = l^*(t,\mathbf{a}) F(N^*, \mathbf{a})$. Next,

$\int_A l^*(t,\mathbf{a}) d\mathbf{a} = q_0(t) \int_A l^*(0,\mathbf{a}) \exp[\sum_{i=1}^{n} \varphi_i(\mathbf{a}) q_i(t)] d\mathbf{a} = N(0) q_0(t) M(0; \mathbf{q}(t)) = N^*(t)$.

This means that the functions $l^*(t,\mathbf{a})$, $N^*(t)$ satisfy system (1)-(2) at $t\in[0,T)$.
Conversely, let $l(t,\mathbf{a})$ and $N(t)$ solve system (1)-(2) for all $t\in[0,T)$, so that

$dl(t,\mathbf{a})/dt = l(t,\mathbf{a}) [f(N) + \sum_{i=1}^{n} g_i(N)\varphi_i(\mathbf{a})], \quad N(t) = \int_A l(t,\mathbf{a}) d\mathbf{a}$.

Let us define for instant the functions $p_i(t)$, $i=0,\ldots n$, by relations:

$p_i(t) = \int_0^t g_i(N(x)) dx$ for $i=1,\ldots m$, $p_0(t) = \exp[\int_0^t f(N(x)) dx]$; then

$dl(t,\mathbf{a})/l(t,\mathbf{a}) = dp_0/p_0 + \sum_{i=1}^{n} \varphi_i(\mathbf{a}) dp_i,$

$\ln l(t,\mathbf{a}) = \ln p_0(t) + \sum_{i=1}^{n} \varphi_i(\mathbf{a}) p_i(t) + C$; taking $C = \ln l(0,\mathbf{a})$, we get

$l(t,\mathbf{a}) = l(0,\mathbf{a}) p_0(t) \exp[\sum_{i=1}^{n} \varphi_i(\mathbf{a}) p_i(t)]$ for all $t\in[0,T)$. Hence

$N(t) \equiv \int_A l(t,\mathbf{a}) d\mathbf{a} = p_0(t) \int_A l(0,\mathbf{a}) \exp[\sum_{i=1}^{n} \varphi_i(\mathbf{a}) p_i(t)] d\mathbf{a} =$

$N(0) p_0(t) M(0; p_1(t),\ldots p_n(t))$ at $t\in[0,T)$. (5.1)
From the definition of $p_i(t)$, $dp_i(t)/dt = g_i(N(t))$ for $i>0$, $dp_0/dt = p_0(t) f(N(t))$
where $N(t)$ is given by (5.1), so $\{p(t)\}$ is a solution of Cauchy problem (4)-(5)
for $t\in[0,T)$. Theorem 1 is proved.

*Proof of Lemma 1.*

$M(t,\lambda) = \{\int_A \exp(\sum_{i=1}^{n} \lambda_i \varphi_i(\mathbf{a})) l(t,\mathbf{a}) d\mathbf{a}\} / N(t) =$

$\{\int_A \exp[\sum_{i=1}^{n} \lambda_i \varphi_i(\mathbf{a})] q_0(t) \exp[\sum_{i=1}^{n} \varphi_i(\mathbf{a}) q_i(t)] l(0,\mathbf{a}) d\mathbf{a}\} / N(t) =$

$$=\{q_0(t) \int_A \exp[\sum_{i=1}^{n} (\lambda_i + q_i(t))\varphi_i(\boldsymbol{a})] \, l(0,\boldsymbol{a})d\boldsymbol{a}\}/ N(t)=$$

$q_0(t)M(0;\boldsymbol{\lambda}+\boldsymbol{q}(t))N(0)/N(t)$.
Further, according to Theorem 1 $N(t) = N(0) \, q_0(t)M(0;\boldsymbol{q}(t))$, so
$M(t,\boldsymbol{\lambda}) = M(0;\boldsymbol{\lambda}+ \boldsymbol{q}(t))/ M(0;\boldsymbol{q}(t))$, as desired.

From the definition of the mgf it is evident, that $E^t\varphi_i(\boldsymbol{a})=\partial_i M(t;0,...\,0)$. Using this relation, we have as a consequence from Lemma 1 an important formula:
$$E^t\varphi_i(\boldsymbol{a})= [\partial_i M(0;\boldsymbol{q}(t))] / M(0;\boldsymbol{q}(t)). \qquad (5.2)$$

*Proof of Theorem 2.*
The first assertion of the theorem follows from Theorem 1. Next,

$$dP(t,\boldsymbol{a})/dt = d\{P(0,\boldsymbol{a}) \exp[\sum_{i=1}^{n} \varphi_i(\boldsymbol{a}) \, q_i(t)] / M(0;\boldsymbol{q}(t))\}/dt =$$

$$P(t,\boldsymbol{a}) [\sum_{i=1}^{n} \varphi_i(\boldsymbol{a}) \, g_i(N(t))] - P(t,\boldsymbol{a}) [\sum_{i=1}^{n} g_i(N(t)) \, \partial_i M(0;\boldsymbol{q}(t))] / M(0;\boldsymbol{q}(t)) =$$

$$P(t,\boldsymbol{a}) [\sum_{i=1}^{n} g_i(N(t))(\varphi_i(\boldsymbol{a}) - E^t\varphi_i)] \, . \text{ Q.E.D.}$$

*Proof of Theorem 3.*
Integrating over $\boldsymbol{a}$ the equation

$$dl(t,\boldsymbol{a})/dt = l(t,\boldsymbol{a}) [f(N)+ \sum_{i=1}^{n} g_i(N)\varphi_i(\boldsymbol{a})]= N(t)P(t,\boldsymbol{a}) [f(N) + \sum_{i=1}^{n} g_i(N)\varphi_i(\boldsymbol{a})],$$

we get (2.4). Formula (2.5) was already proved (see (5.2)). Next,

$$d(E^t\varphi_i)/dt = d[\int_A \varphi_i(\boldsymbol{a})P(t,\boldsymbol{a})d\boldsymbol{a}]/dt = \text{(by Theorem 2)}$$

$$\int_A \varphi_i(\boldsymbol{a})[\sum_{k=1}^{n} g_k(N(t))(\varphi_k(\boldsymbol{a}) - E^t\varphi_k)] \, P(t,\boldsymbol{a})d\boldsymbol{a} =$$

$$\sum_{k=1}^{n} g_k(N(t))\{E^t[\varphi_k(\boldsymbol{a})\varphi_i(\boldsymbol{a})] -(E^t\varphi_k)(E^t\varphi_i)\}= \sum_{k=1}^{n} g_k(N)K(t;i,k).$$

Formula (2.6) easily follows from the definition of mgf $M(t;\boldsymbol{\lambda})$ and Lemma 1.

*Proof of Theorem 4*
Let us prove the last assertion (iii). $\Gamma$- distribution (2.9) has the mean $\mu = \eta + k/s$, the variance $\sigma^2 = k/s^2$ and the mgf $M(\lambda) = \exp(\lambda \eta) /(1-\lambda/s)^k$ for $\lambda<s$. According to Lemma 1, $M(t;\lambda) = M(0;\lambda+q_1(t))/M(0;q_1(t)) =$
$\exp(\lambda b) \, [(s - \lambda - q_1(t))/( s - q_1(t))]^{-k} = \exp(\lambda b) \, [1- \lambda /( s - q_1(t))]^{-k}$.
It means that the current distribution $P(t,a)$ at $t<T^*$ is $\Gamma$-distribution (2.9) with coefficients $(s- q_1(t))$, $k$, $b$. The mean and the variance of this distribution are given by (2.8), (2.9). Other assertions of the theorem can be proved similarly.

*Proof of Theorem 5*

By conditions, $M(0;\pmb{\lambda})=\prod_{i=1}^{n} M_i(0;\lambda_i)$; then according to Lemma 1

$M(t,\pmb{\lambda}) = M(0;\pmb{\lambda}+ \pmb{q}(t))/ M(0;\pmb{q}(t))= \prod_{i=1}^{n} M_i(0;\lambda_i+q_i(t))/M_i(0;q_i(t))$. This mgf corresponds to the distribution $\pmb{p}(t,x_1, \ldots x_n)=\prod_{i=1}^{n} p_i(t,x_i)$ of a random $n$-vector with independent components, where $p_i(t,x_i)$ is a pdf of a r.v. with the mgf $M_i(0;\lambda_i+q_i(t))/M_i(0;q_i(t))$. As the mgf uniquely determines the distribution, the proof is completed.

*Proof of Theorem 6*

Let us prove the assertion iii). Due to Lemma 1, $M(t;\pmb{\lambda}) = M(0;\pmb{\lambda}+\pmb{q}(t))/M(0;\pmb{q}(t))=$ $\exp[s(\theta+\lambda+q(t))-s(\theta)]/\exp[s(\theta+q(t))-s(\theta)]=\exp[s(\theta+\lambda+q(t))- s(\theta+q(t))]$, as desired. All other assertions of the theorem can be proved the same way.

**Conclusion.** The main problem of the dynamic theory of inhomogeneous populations is how the heterogeneity of a population affects its dynamics. The theory described here predicts several essential, new dynamic regimes applicable even to well-known, simple population models. In the general case, neither the average fitness nor its variance is sufficient to predict the overall population growth; instead, the entire initial distribution of the fitness over the individuals should be used to investigate real population dynamics. We developed methods for effective analysis of heterogeneous models, which allow us to explore the dynamics of the total population size together with the current distribution of the vector-parameter. We showed that, typically, there exists a phase of "hyper-exponential" growth (similar to that predicted by the Fisher's fundamental theorem of natural selection) which precedes the well-known exponential phase of population growth in a free regime. The developed formalism was successfully applied to various problems of mathematical biology, such as inhomogeneous Malthusian and logistic models (Karev, 2000, 2005), the evolution of trait distributions under natural selection (Cunningham et al, 2001), tree stand self-thinning models (Karev, 2003), prey-predator models (Novozhilov, 2004), models of global demography (Karev, 2005). Investigation of complex population models (like inhomogeneous Allee-type model) and applications to general selection theory (G. Price, 1995) are promising next steps.

**Acknowledgement.** Author thanks Dr. E. Koonin and Dr. A. Novozhilov for valuable discussions and help in preparation of the manuscript.